\documentclass[twocolumn]{jpsj3} 
\usepackage{bm,amssymb,amsmath}
\usepackage{graphicx}   
\usepackage{txfonts}
\newcommand{\simg}{\stackrel{>}{_\sim}}
\newcommand{\siml}{\stackrel{<}{_\sim}}

\title{
FFLO Superconductivity Mediated by Excitonic Fluctuation \\in Semimetallic Ta$_2$NiSe$_5$
}
\author{Takemi YAMADA$^1$\thanks{E-mail address: t-yamada@rs.tus.ac.jp},
Kaoru DOMON$^2$, 
Yoshiaki \=ONO$^2$
}
\inst{
$^1$Department of Physics, Faculty of Science and Technology, Tokyo University of Science, Chiba, 278-8510, Japan \\
$^2$Department of Physics, Niigata University, Niigata, 950-2181, Japan
}
\abst{
We investigate the excitonic fluctuation and its mediated superconductivity in the quasi one-dimensional three-chain Hubbard model for Ta$_2$NiSe$_5$ known as a candidate material for the excitonic insulator.
In the semiconducting case and the semimetallic case with a small band-overlapping where one conduction ($c$) band and one valence ($f$) band cross the Fermi level, the excitonic fluctuation with $\bm{q}=\bm{0}$ is enhanced due to the $c$-$f$ Coulomb interaction and diverges towards the uniform excitonic order corresponding to the excitonic insulator.
On the other hands, in the semimetallic case with a large band-overlapping where two $c$ bands and one $f$ band cross the Fermi level, the non-uniform excitonic fluctuation with $\bm{q}\neq \bm{0}$ corresponding to the nesting vector between the $c$ and $f$ Fermi-surfaces (FSs) becomes dominant and results in the Fulde-Ferrell-Larkin-Ovchinnikov (FFLO) excitonic order characterized by the condensation of excitons with finite center-of-mass momentum $\bm{q}$.
Near the instability, the largely enhanced excitonic fluctuations mediate the $c$-$f$ interband Cooper pairs with finite center-of-mass momentum resulting in the FFLO superconductivity, which is expected to be realized in the semimetallic Ta$_2$NiSe$_5$ under high pressure.
}

\begin{document}
\maketitle

\section{Introduction}\label{sec1}
Excitonic condensation in the electron-hole (e-h) system has been a central issue in the condensed matter physics since about six decades ago\cite{Knox.1963,PR.158.462,RMP.40.755}, 
where conduction electrons and valence holes can form bound states, and they spontaneously drive the system into the low-temperature ground state, called the excitonic phase (EP) or excitonic insulator (EI). 
Recent discoveries of new candidate materials 
gather 
a growing interest of EP in the real materials in terms of 
the emergence of the BCS-Bose-Einstein condensation (BEC) crossover 
\cite{PRB.74.165107,PRB.78.193103,PRB.81.205117,PRB.84.245106,PRB.85.121102,JPCM.27.333201}. 

Recently, the ternary chalcogenide Ta$_2$NiSe$_5$ has received renewed attention as a strong candidate for the EI. 
The resistivity indicates a narrow gap semiconductor with a quasi one-dimensional (1-D) structure where a Ni chain and adjacent two Ta chains construct the 1-D three chain\cite{Inorg.Chem.24.3611,JLCM.116.51}. A second-order structural transition from the orthorhombic to monoclinic phase occurs at a critical temperature $T_S$=328 K\cite{JLCM.116.51}, below which a remarkable flattening of the valence band top is observed in the angular resolved photo emission spectroscopy (ARPES) 
experiments\cite{PRL.103.026402,JSNM.25.1231} without any sign of the magnetic and density-wave-type order. 

Several theoretical studies have revealed that the transition is well accounted for by the excitonic condensation from a normal semiconductor (orthorhombic) to the EI (monoclinic) from a mean-field (MF) analysis for the 1-D three-chain Hubbard model\cite{PRB.87.035121,PRB.93.041105} and from a variational cluster approximation for the extended Falicov-Kimball model\cite{PRB.90.155116}. 
Recent optical measurements are also consistent with the EI phase below $T_S$ \cite{Nat.Commun.8.14408,Larkin2017} 
and is analyzed by the band structure calculation together with the density-matrix renormalization group study, 
exhibiting the strong coupling EI with the BEC of the exciton-pairs.\cite{PRL.120.247602}

When the pressure is applied for Ta$_2$NiSe$_5$\cite{Nakano2018,Matsubayashi}, $T_S$ is suppressed and the system changes from semiconducting to semimetallic both above and below $T_S$, and then, $T_S$ finally becomes zero at a critical pressure $P_c\sim 8$ GPa, around which the superconductivity is observed. 
Then, we have recently investigated the 1-D three-chain Hubbard model also in the semimetallic case and have found that the difference of the band degeneracy, the two-fold degenerate conduction bands and the nondegenerate valence band, inevitably causes the imbalance of each Fermi wavenumber and results in a remarkable excitonic state characterized by the condensation of excitons with finite center-of-mass (COM) momentum $q$, the so-called Fulde-Ferrell-Larkin-Ovchinnikov (FFLO) excitonic state\cite{JPSJ.85.053703,JPSJ.85.065005}, as previously discussed in the electron-hole bilayer systems with density imbalance\cite{PRB.75.113301,JPSJ.79.033001,PRB.81.115329}. More recently, a realistic quasi 1-D three-chain Hubbard model\cite{PRB.94.085111} has also been studied and a detailed excitonic phase diagram including the FFLO state has been obtained\cite{JPSJ.87.054701}.

As the superconductivity is observed near the excitonic phase in Ta$_2$NiSe$_5$, the excitonic fluctuation is considered to be a key ingredient for the pairing mechanism as early discussed by Little for quasi 1-D organic superconductors\cite{Little} but has not been discussed for Ta$_2$NiSe$_5$ so far. The purpose of this paper is to investigate the enhancement of the excitonic fluctuations in quasi 1-D semimmetallic situation and superconductivity mediated by such the excitonic fluctuations in Ta$_2$NiSe$_5$ on the basis of the quasi 1-D three-chain Hubbard model where the FFLO excitonic order is found to take place in the semimetallic case\cite{JPSJ.87.054701}.

This paper is organized as follows: In Sec. \ref{sec2}, 
we introduce a model Hamiltonian and give the formulation of the random phase approximation (RPA) and Eliashberg equation for the present model. 
Next in Sec. \ref{sec3}, we show the results for the normal-excitonic phase boundary diagram 
and the enhancement of the excitonic fluctuation and its effect on the superconductivity. 
In Sec. \ref{sec4} we give the summary and discussion about the present mechanism and experimental prediction. 

\section{Model \& Formulation}\label{sec2}
\subsection{Hamiltonian}
\begin{figure}[t]
\centering
\includegraphics[width=8.7cm]{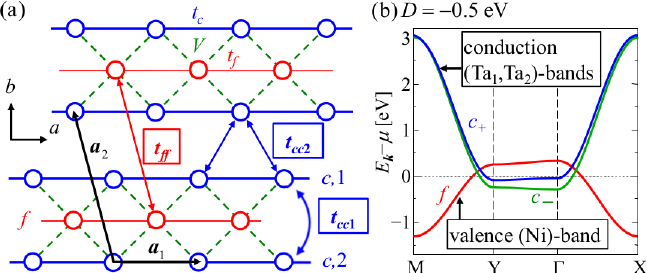}
\caption{(Color online) 
(a) Schematic representation of the quasi 1-D three-chain Hubbard model for Ta$_2$NiSe$_5$  where $\bm{a}_1$ and $\bm{a}_2$ are unit lattice vectors. 
(b) Band dispersion along the high-symmetry line M[$(\pi.\pi)$]$-$Y[$(0,\pi)$]$-$$\Gamma$[$(0,0)$]$-$X[$(\pi,0)$].
}
\label{Fig1}
\end{figure}
The quasi 1-D three-chain Hubbard model for Ta$_2$NiSe$_5$\cite{PRB.93.041105,JPSJ.87.054701} is schematically shown in Fig. \ref{Fig1} (a), where a Ni ($f$) chain and adjacent two Ta ($c$) chains construct the 1-D three chains\cite{PRB.87.035121}, 
and these chains are weakly coupled with the neighboring three chains, which is explicitly given by the Hamiltonian, 
\begin{align}
H&=H_{0}+H_{V},\label{eq:H}\\
&H_{0}=\sum_{\bm{k}\sigma} \Psi_{\bm{k}\sigma}^{\dagger}\hat{H}_{0}(\bm{k})\Psi_{\bm{k}\sigma},\\
&\hat{H}_{0}(\bm{k})=
\left(
\begin{array}{ccc}
\!\varepsilon_{\bm{k}c}-\mu & \varepsilon_{\bm{k}c}\rq{} & 0 \!\\
\!\varepsilon_{\bm{k}c}\rq{}^{*} & \varepsilon_{\bm{k}c}-\mu & 0 \!\\
\!0 & 0 &\varepsilon_{\bm{k}f}-\mu \!\\
\end{array}
\right),\\
&H_{V}=V\sum_{\bm{i}\alpha}\left(n^{c}_{\bm{i}\alpha}+n^{c}_{\bm{i}-\bm{a}_1\alpha}\right)n^{f}_{\bm{i}},
\end{align}
where $\Psi_{\bm{k}\sigma}^{\dagger}=\left(c^{\dagger}_{\bm{k}1\sigma},c^{\dagger}_{\bm{k}2\sigma},f^{\dagger}_{\bm{k}\sigma}\right)$ and $c_{\bm{k}\alpha\sigma}(c_{\bm{i}\alpha\sigma})$, $f_{\bm{k}\sigma}(f_{\bm{i}\sigma})$ are the annihilation operators for the $c$, $f$ electrons with wavevector $\bm{k}$ (unit cell at $\bm{i}=n_{1}\bm{a}_1+n_{2}\bm{a}_2$ with unit lattice vectors $\bm{a}_1$ and $\bm{a}_2$), spin $\sigma=\uparrow,\downarrow$ and chain index for the $c$ electron $\alpha=1,2$.  $n_{\bm{i}\alpha}^{c}=\sum_{\sigma}c^{\dagger}_{\bm{i}\alpha\sigma}c_{\bm{i}\alpha\sigma}$,  $n_{\bm{i}}^{f}=\sum_{\sigma}f^{\dagger}_{\bm{i}\sigma}f_{\bm{i}\sigma}$ are the number operators and $\mu$ is a chemical potential. 
The matrix elements of $\hat{H}_{0}(\bm{k})$ are given by, 
\begin{align}
&\varepsilon_{\bm{k}c}=D/2+2t_{c}\left({\rm cos}\bm{k}\cdot\bm{a}_1-1\right),\\
&\varepsilon_{\bm{k}c}\rq{}=t_{cc1}+2t_{cc2}\left(e^{i\bm{k}\cdot(\bm{a}_1+\bm{a}_2)}+e^{i\bm{k}\cdot\bm{a}_2}\right),\\
&\varepsilon_{\bm{k}f}=-D/2+2t_{f}\left({\rm cos}\bm{k}\cdot\bm{a}_1-1\right)\nonumber\\
&\qquad+2t_{ff}\left({\rm cos}\bm{k}\cdot(\bm{a}_1+\bm{a}_2)+{\rm cos}\bm{k}\cdot\bm{a}_2-2\right),
\end{align}
where the intra-chain hopping parameters, $t_{c}$ and $t_{f}$, 
and the additional inter-chain hopping parameters, $t_{cc1}$, $t_{cc2}$ and $t_{ff}$, 
are set to $t_{c}=-0.8$ eV, $t_{f}=0.4$ eV, $t_{cc1}=-0.02$ eV,  $t_{cc2}=-0.05$ eV, and $t_{ff}=0.01$ eV, 
which are the same with the values of the previous study.\cite{JPSJ.87.054701}
The resulting conduction bonding (anti-bonding) $c_{\mp}$ band, 
\begin{align}
\epsilon_{\bm{k}c\mp}=\frac{\epsilon_{\bm{k}c}\mp\left|\epsilon_{\bm{k}c}\rq{}\right|}{2}
\end{align}
and the valence $f$ band $\epsilon_{\bm{k}f}$ are plotted in Fig. \ref{Fig1} (b). 
$D$ is the band-overlap or band-gap where 
the $c_{-}$ and $f$ bands meet just at $\bm{k}=\bm{0}$ when $D=D_{0}(=0.12$ eV) which gives the boundary between the semiconducting ($D>D_0$) and semimetallic ($D<D_0$) regime. 
As $D$ is considered to be a decreasing function of pressure, 
we vary $D$ as a parameter to examine the pressure effect on this model from semiconducting to semimetallic regime. 
In fact, we have estimated the band-overlap $|D|$ from the first-principles band calculation under pressure and have found that $|D|$ increases with increasing the pressure although the available lattice parameters 
under the pressure are restricted\cite{Nat.Commun.8.14408} and is difficult to obtain the explicit pressure dependence of $D$.

In Eq. (\ref{eq:H}), we consider the intersite $c$-$f$ Coulomb interaction $V$ which is crucial for the excitonic order and its fluctuation as shown below, while we neglect the on-site Coulomb interaction $U$ which can be effectively included in $D$ and/or the chemical potential $\mu$ within the MF approximation by excluding the magnetic and density-wave-type orders\cite{PRB.87.035121,PRB.93.041105,JPSJ.85.053703,JPSJ.85.065005,PRB.94.085111,JPSJ.87.054701}. In fact, such orders have not been observed experimentally in Ta$_2$NiSe$_5$. 

\subsection{Random Phase Approximation}
\begin{figure}[t]
\centering
\includegraphics[width=8.7cm]{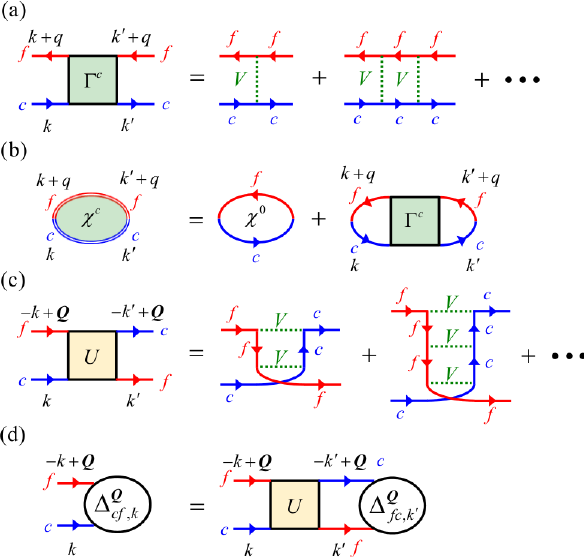}
\caption{(Color online) 
Diagrammatic representation of (a) the charge-orbital vertex with a infinite summation of ladder diagram with respect to $V$ ($c$-$f$ excitonic mode) $\Gamma_{cfcf}^{c}(\bm{k},\bm{k}\rq{},\bm{q})$ and of (b) the excitonic susceptibility $\chi_{cfcf}(k\!,k\rq{}\!,q)$, 
and of (c) The effective pairing interaction $U(k,k',\bm{Q})$ between the $c$ and $f$ electrons mediated by the excitonic fluctuation and of (d) the linearized Eliashberg equation for superconducting gap function including the FFLO solution with the finite COM momentum $\bm{Q}$.
}
\label{Fig2}
\end{figure}
Here we employ the RPA which can describe 
the enhancement of the fluctuations around the MF critical point of the excitonic ordered phase 
recently obtained in the our previous work\cite{JPSJ.87.054701}. 
Within the RPA, the effect of the excitonic fluctuation is described by a summation of the ladder diagrams with respect to $V$ 
for the charge-orbital vertex $\hat{\Gamma}^{c}(k\!,k\rq{}\!,q)$ 
as shown in Fig. \ref{Fig2} (a) which is given by the $9\times9$ matrix representation as, 
\begin{align}
\hat{\Gamma}^{c}(k\!,k\rq{}\!,q)\!\!&=\!\!\hat{\Gamma}^{0}(\bm{k}\!,\bm{k}\rq{}\!)
\nonumber\\
&\quad\quad
\!+\!\sum_{k\rq{}\rq{}}\hat{\Gamma}^{0}(\bm{k}\!,\bm{k}\rq{}\rq{}\!)\hat{\chi}^{0}(k\rq{}\rq{}\!,q)\hat{\Gamma}^{c}(k\rq{}\rq{}\!,\!k\rq{}\!,q),\label{eq:vtx}\\
\hat{\chi}^0(k\!,q)\!\!&=\!\!-\hat{G}^0(k+q)\hat{G}^0(k),\\
\hat{G}^0(k)&=\left[i\varepsilon_n\hat{1}-\hat{H}_0(\bm{k})\right]^{-1},
\end{align}
where $\hat{\chi}^0(k\!,q)$ and $\hat{G}^0(k)$ are the particle-hole bare susceptibility and 
the Green's function ($3\times3$ matrix) with $q=(\bm{q},i\omega_{\nu})$, $k=(\bm{k},i\varepsilon_n)$ and the momentum $\bm{k}, \bm{q}$, 
and the fermionic and bosonic Matsubara frequencies $\varepsilon_{n}=(2n+1)\pi T$, $\omega_{\nu}=2\nu\pi T$, respectively.
$\ell$ is the site-orbital channels $\ell$=1,2,3 (=$c_1,c_2,f$) and 
$\hat{1}$ is the unit matrix in the site-orbital basis. 

The charge-orbital bare vertex due to the intersite Coulomb $V$ is 
\begin{align}
&[\Gamma^{0}(\bm{k},\bm{k}\rq{})]_{\ell_1\ell_2,\ell_3\ell_4}\nonumber\\
&\!=\!\left\{
\begin{array}{ll}
\!V(1+e^{(\bm{k}-\bm{k}\rq{})\cdot\bm{a}_1}) & \ell_1\!=\!\ell_3\!=\!\alpha(\!f\!), \ell_2\!=\!\ell_4\!=\!f(\!\alpha\!), \\
\!0 & {\rm otherwise},\\
\end{array}
\!
\right.
\end{align}
where we neglect the magnetic and density-wave-type fluctuations due to the effect of $U$ together with the ring diagrams with respect to $V$ since the same reason mentioned in the previous subsection. 
We also discuss the possibility of such the density, hybridization, and pair-formation-type condensation
within the MF approximation for the intersite $c$-$f$ Coulomb interaction in the present three-chain model
in Appendix \ref{sec-V}.

In order to take the effect of the inter-cell $c$-$f$ Coulomb interaction in the susceptibility, 
we transform $\hat{\Gamma}^0(\bm{k}\!,\bm{k}\rq{}\!)$ and $\hat{\chi}^0(k\!,q)$ in Eq. (\ref{eq:vtx}) into the $13\times13$ matrix representations $\widetilde{\Gamma}^0_{\alpha\beta}$ and $\widetilde{\chi}^0_{\alpha\beta}$ with $\alpha,\beta=1\sim13$, respectively, 
\begin{align}
&[\Gamma^{0}(\bm{k}\!,\bm{k}\rq{}\!)]_{\ell\ell\rq{},mm\rq{}}
=\sum_{\alpha\beta}g_{\ell\ell\rq{}}^{\alpha}(\bm{k})~\widetilde{\Gamma}^{0}_{\alpha\beta}~g^{\beta *}_{mm\rq{}}(\bm{k}),\\
&\widetilde{\chi}^0_{\alpha\beta}(q)\!=\!\sum_{k}\!\!
\sum_{\ell\ell\rq{}mm\rq{}}g_{\ell\ell\rq{}}^{\alpha*}(\bm{k})
[\hat{\chi}^0(k,q)]_{\ell\ell\rq{},mm\rq{}}
g^{\beta}_{mm\rq{}}(\bm{k}),
\end{align}
where $g_{\ell\ell\rq{}}^{\alpha}(\bm{k})$ is the extended basis function similar to the three band $d$-$p$ model for high-$T_c$ cuprate\cite{PRB.39.12371}
and the explicit form is given in Appendix \ref{sec-ebf}. 
Then we obtain the charge-orbital susceptibility 
as shown in Fig. \ref{Fig2} (b), 
\begin{align}
&\hat{\chi}^{c}(q)=\sum_{kk\rq{}}\hat{\chi}^{c}(k\!,k\rq{}\!,q),\\
&\hat{\chi}^{c}(k\!,k\rq{}\!,q)=\hat{\chi}^{0}(k\!,q)\left(\delta_{kk\rq{}}
+\hat{\Gamma}^c(k\!,k\rq{}\!,q)\hat{\chi}^{0}(k\rq{}\!,q)\right), \\
&[\hat{\Gamma}^c(k\!,k\rq{}\!,q)]_{\ell\ell\rq{},mm\rq{}}
=\sum_{\alpha\beta}g_{\ell\ell\rq{}}^{\alpha}(\bm{k})~\widetilde{\Gamma}^{c}_{\alpha\beta}(q)~g^{\beta *}_{mm\rq{}}(\bm{k}),
\end{align}
where 
$\widetilde{\Gamma}^{c}_{\alpha\beta}(q)=[\widehat{\Gamma}^{c}(q)]_{\alpha\beta}$ and 
$\widetilde{\chi}^{0}_{\alpha\beta}(q)=[\widehat{\chi}^{0}(q)]_{\alpha\beta}$, and 
\begin{align}
\widehat{\Gamma}^{c}(q)=\left[\widehat{1}-\widehat{\Gamma}^{0}\widehat{\chi}^{0}(q)\right]^{-1}\widehat{\Gamma}^{0}.
\end{align} 
When the largest eigenvalue of $\widehat{\Gamma}^{0}\widehat{\chi}^{0}(\bm{q},\!i\omega_{\nu}\!=\!0)$ reaches unity, the charge-orbital instability including the excitonic one with the ordering momentum $\bm{q}$ takes place. 
The more explicit formulation of the phase boundary condition within the MF approximation 
for the excitonic phase order is given in Appendix \ref{sec-pbc}. 


\subsection{Linearized Eliashberg equation \& Effective pairing interaction}
In order to examine the superconductivity in the $c$-$f$ density imbalanced systems, 
we solve the linearized Eliashberg equation with including the FFLO solution shown in Fig. \ref{Fig2} (d) which is given by 
\begin{align}
&\lambda^{\bm{Q}} \Delta_{\ell\ell\rq{}}^{\bm{Q}}(k)
\!\!=\!\!-
\sum_{k\rq{}}\sum_{\ell_1\ell_2\ell_3\ell_4} [\hat{U}(k\!,k\rq{}\!,\bm{Q})]_{\ell \ell_1,\ell_2 \ell\rq{}} \nonumber \\
&\qquad\quad\times\!\![\hat{G}^0(-k\rq{}\!+\!\bm{Q})]_{\ell_3 \ell_1}\Delta_{\ell_3 \ell_4}^{\bm{Q}}(k\rq{})
[\hat{G}^0(k\rq{})]_{\ell_4 \ell_2}, 
\label{eq:gap}
\end{align}
where $\Delta_{\ell\ell\rq{}}^{\bm{Q}}(k)$ is the superconducting (SC) gap function with the COM momentum of Cooper pairs $\bm{Q}$. $\hat{U}(k,k\rq{},\bm{Q})$ is the $9\times9$ matrix representation of the effective pairing interaction for the spin-singlet state mediated by the charge-orbital (excitonic) fluctuation diagrammatically shown in Fig. \ref{Fig2} (c) and explicitly given by
\begin{align}
&\hat{U}(k\!,k\rq{}\!,\bm{Q})\nonumber\\
&\!=\!\!-\!\frac{1}{2}
\sum_{k_1k_2}\!
\hat{\Gamma}^{0}(\bm{k},\bm{k}_1)
\hat{\chi}^{c}(k_1,\!-\!k_2\!+\!\bm{Q},k_1\!-\!k_2)\hat{\Gamma}^{0}(-\bm{k}_2\!+\!\bm{Q},\bm{k}\rq{}) 
\nonumber\\
&\!=\!-\frac{1}{2}[\hat{\Gamma}^c(k,\!-\!k\rq{}\!+\!\bm{Q},k\!-\!k')\!-\!\hat{\Gamma}^0(\bm{k}\!,\bm{k}\rq{})],
\end{align}
which 
yields a large attractive interaction for $\bm{k}-\bm{k'}\sim \bm{q}$ near the charge-orbital instability with $\bm{q}$ almost independent of $\bm{Q}$. Eq. (\ref{eq:gap}) is solved to obtain the gap function $\Delta_{\ell\ell\rq{}}^{\bm{Q}}(k)$ with the eigenvalue $\lambda^{\bm{Q}}$ which becomes unity at the SC transition temperature $T_c$. 

Throughout the paper, the chemical potential $\mu$ is set so as to keep the electron number per unit cell: $\langle n\rangle=\sum_{\alpha}\langle n_{\bm{i}\alpha}^{c}\rangle+\langle n_{\bm{i}}^{f}\rangle=2$ 
for any given band-overlap $D$ which corresponds to the situation of the semiconducting and semimetallic cases in Ta$_2$NiSe$_5$. 
The RPA and SC calculations are performed 
with 1024$\times$16 $\bm{k}$-point meshes and 2048 Matsubara frequencies 
and $V=0.6$ eV as a realistic value of the case with weak screening. Here and hereafter, we measure the energy in units of eV.

\section{Results}\label{sec3}
Here we focus only on the phase boundary owing to consider the excitonic fluctuation and the superconductivity
and thus we set aside the electronic states below $T_{\rm EP}$ which is precisely studied in Ref.\cite{JPSJ.87.054701}

\begin{figure}[t]
\centering
\includegraphics[width=8.7cm]{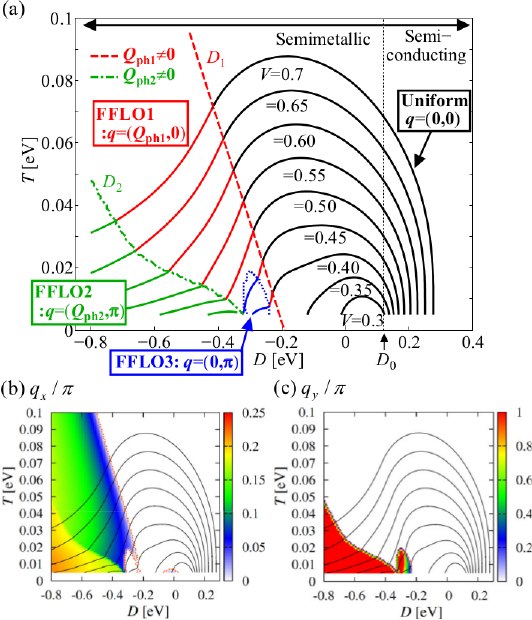}
\caption{(Color online) 
(a) $D$-$T$ phase boundaries between normal-excitonic phases for several values of $V>0$. 
The color maps for the values of $q_x/\pi$ (b) and $q_y/\pi$ (c) on the $D$-$T$ phase diagram. 
}
\label{Fig3}
\end{figure}
\begin{figure}[t]
\centering
\includegraphics[width=8.7cm]{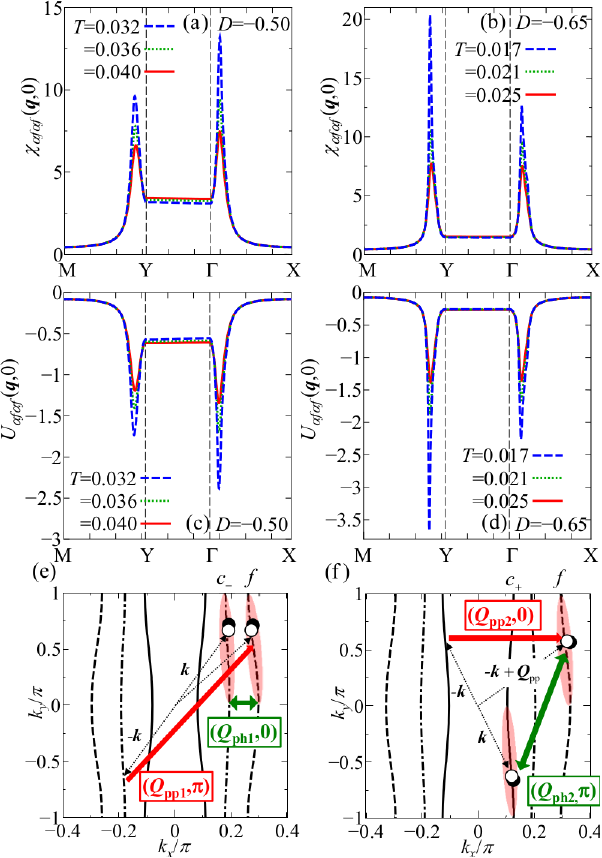}
\caption{(Color online) 
Excitonic component of the charge-orbital susceptibility $[\hat{\chi}^c(\bm{q},0)]_{\alpha f,\alpha f}$ with $\alpha=c_1,c_2$ at several temperatures for $D=-0.5$ (a) and $-0.65$ (b) along the high-symmetry line, and the corresponding pairing interaction $[\hat{U}(\bm{q},0)]_{\alpha f,\alpha f}$ with $\alpha=c_1,c_2$ for $D=-0.5$ (c) and $-0.65$ (d). 
The FSs of $c_+$ (solid line), $c_-$ (dot-dashed line) and $f$ (dashed line) bands for $D=-0.5$ (e) and $-0.65$ (f) in the narrow region of $-0.4 \leq k_x/\pi \leq 0.4$, where the nesting vector of the particle-hole channnel $(Q_{\rm ph1},0)$ ($(Q_{\rm ph2},\pi)$) together with that of the particle-particle channel $(Q_{\rm pp1},\pi)$ ($(Q_{\rm pp2},0)$) between $f$ and $c_-$ ($c_+$) FSs is plotted in (e) ((f)). 
}
\label{Fig4}
\end{figure}
First we show the normal-excitonic phase (EP) transition temperature $T_{\rm EP}$ within the RPA as functions of $(V\!,\!T\!,\!D)$ as shown in Fig. \ref{Fig3} (a) where we search the maximum value of the bare excitonic susceptibility $\chi_{\rm ex}^{0}$($\bm{q},i\omega_{\nu}$=$0)$ and determine the condition $V\chi_{\rm ex}^{0}(\bm{q},0)$=$1$ for any given $(V\!,\!T\!,\!D)$ together with the COM momentum $\bm{q}$, which yields the critical values of $(V\!,\!T\!,\!D)$ [See Appendix \ref{sec-pbc}]. 
As well as the previous 1-D model study\cite{JPSJ.85.053703,JPSJ.85.065005}, 
the dome-like structure of $T_{\rm EP}$ between semiconducting $(D>D_{0})$ and semimetallic $(D<D_{0})$ region 
is obtained depending on $T$ and $V$, 
which originates from the van-Hove singularity of the density of states (DOS) in the present quasi 1-D nature 
together with the thermal effect on the excitonic susceptibility $\chi_{\rm ex}^{0}(\bm{q},0)$
which explicitly includes the Fermi distribution function and satisfies the condition $V\chi_{\rm ex}^{0}(\bm{q},0)=1$ at finite $T$ [See Appendix \ref{sec-pbc}]. 
In contrast to the 1-D study\cite{JPSJ.85.053703,JPSJ.85.065005}, 
the system enters into four distinct EPs; a conventional uniform EP with zero COM wavevector $\bm{q}=\bm{0}$, and three FFLO EPs, FFLO1, FFLO2 and FFLO3 with finite $\bm{q}=(q_x,q_y)$ 
as plotted in color maps of $q_x$ [Fig. \ref{Fig3} (b)] and $q_y$ [Fig. \ref{Fig3} (c)] on the $D$-$T$ phase diagrams. 

The uniform EP is stabilized over $D>D_0$ and $D<D_0$ with a small $|D|$, 
and in particular below $D<D_0$, the system becomes the semimetallic with a single $c_{-}$-band and single $f$-band crossing the Fermi level 
where the conventional (uniform) exciton condensation is achieved.
On the other hands, 
with decreasing $D$ below $D_1$ where 
a new Fermi-surface (FS) of the $c_{+}$-band emerges, 
the $c$-$f$ density imbalance is inevitably induced as well as the e-h bilayer systems\cite{PRB.75.113301,JPSJ.79.033001,PRB.81.115329}. 
In such the situation, 
the FFLO1 exicitonic fluctuation given by the maximum of $\chi_{\rm ex}^{0}(\bm{q},0)$ with the nesting wavevector between  $c_{-}$-and $f$-bands, \lq\lq{}$c_{-}$-$f$ band nesting\rq{}\rq{} $\bm{q}=(Q_{\rm ph1},0)$, becomes more dominant than that of uniform one. 
Furthermore with decreasing $D$ and $T$, a new FFLO fluctuation FFLO2 with the \lq\lq{}$c_{+}$-$f$ band nesting\rq{}\rq{} wavevector $\bm{q}=(Q_{\rm ph2},\pi)$ dominates over the FFLO1 around $D<D_2$. 
A third fluctuation FFLO3 with $\bm{q}=(0,q_y) $[$q_y=0\sim\pi$] is also observed in lower $T$ between the two FFLO fluctuations. 
As discussed in the MF study\cite{JPSJ.87.054701}, 
The weak $q_y$-dependent peak structure of $\chi_{\rm ex}^{0}(\bm{q},0)$ along $\bm{q}=(Q_{\rm ph1},0)\sim(Q_{\rm ph2},\pi)$
is the origin of the present several FFLO phases as explicitly shown in Appendix \ref{sec-qdep}. 

Figures \ref{Fig4} (a) and (b) show the static charge-orbital susceptibility $[\hat{\chi}^c(\bm{q},0)]_{\alpha f,\alpha f}$ with $\alpha=c_1,c_2$ corresponding to the excitonic fluctuation at several temperatures for $D=-0.5$ and $-0.65$. 
There are two sharp peaks at $\bm{q}=(Q_{\rm ph1},0)$ and $(Q_{\rm ph2},\pi)$ corresponding to the particle-hole nesting vectors between the $f$ and $c_{-}$ FSs and the $f$ and $c_{+}$ FSs as shown in Figs. \ref{Fig4} (e) and (f), respectively, where the former (latter) peak dominates over the latter (former) one for $D=-0.5$ ($D=-0.65$). 
When $T$ decreases, $[\hat{\chi}^c(\bm{q},0)]_{\alpha f,\alpha f}$ increases and finally diverges with $\bm{q}=(Q_{\rm ph1},0)$ ($(Q_{\rm ph2},\pi)$) for $D=-0.5$ ($D=-0.65$) at a critical temperature $T_{\rm EP}$ below which the FFLO1 (FFLO2) excitonic order with the finite COM momentum of excitons $(Q_{\rm ph1},0)$ ($(Q_{\rm ph2},\pi)$) takes place as previously shown in our previous paper on the basis of the MF analysis\cite{JPSJ.87.054701}. 

\begin{figure}[t]
\centering
\includegraphics[width=8.7cm]{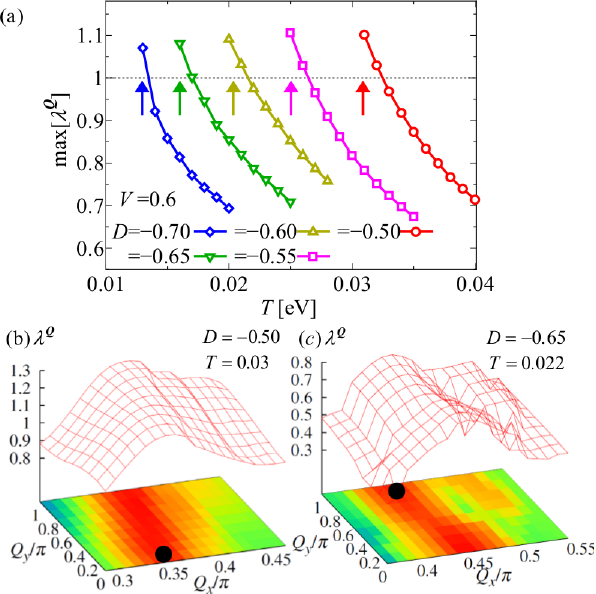}
\caption{(Color online) 
Maximum values of the eigenvalue $\lambda^{\bm{Q}}$ of the Eliashberg equation for several values of $D$ as functions of $T$ (a) where arrows show $T_{\rm EP}$ for correspondong $D$, and the ${\bm{Q}}$-dependence of $\lambda^{\bm{Q}}$ for $D=-0.5$ (b) and -0.65 (c) where $\lambda^{\bm{Q}}$ is maximum at ${\bm{Q}}$ with the closed circle. 
}
\label{Fig5}
\end{figure}

When $T$ approaches $T_{\rm EP}$, the effective pairing interaction between the $c$ and $f$ electrons $[\hat{U}(k,k',\bm{Q})]_{\alpha f,\alpha f}$ with $\alpha=c_1,c_2$ is largely enhanced for $\bm{k}-\bm{k'}\sim(Q_{\rm ph1},0)$ and $(Q_{\rm ph2},\pi)$ due to the largely enhanced excitonic fluctuation mentioned above, and then mediates the $c$-$f$ pairing with the gap function $\Delta_{\alpha f}^{\bm{Q}}(k)$. 
The explicit results of the pairing interactions $[\hat{U}(\bm{q},0)]_{\alpha f,\alpha f}\equiv [\hat{U}(k,k',0)]_{\alpha f,\alpha f}$ with $\bm{q}$=$\bm{k}-\bm{k'}$ and $\alpha=c_1,c_2$ are plotted in Fig. \ref{Fig4} (c) and (d) as the same plot of the corresponding $[\hat{\chi}^c(\bm{q},0)]_{\alpha f,\alpha f}$ as shown in Fig. \ref{Fig4} (a) and (b), where $[\hat{U}(\bm{q},0)]_{\alpha f,\alpha f}$ is largely developed with the corresponding wavevector $\bm{q}\sim(Q_{\rm ph1},0)$ and $(Q_{\rm ph2},\pi)$ as a attractive pairing interaction between $\Delta_{\alpha f}^{\bm{Q}}(k)$. 

In Fig. \ref{Fig5} (a), we plot the $T$-dependence of the eigenvalue $\lambda^{\bm{Q}}$ of the Eliashberg equation Eq. (\ref{eq:gap}) for several values of $D$, where we set the value of ${\bm{Q}}$ for which $\lambda^{\bm{Q}}$ becomes maximum. Systematic calculations of $\lambda^{\bm{Q}}$ for various $D$ reveal that, for $D=-0.5$ and $-0.55$ ($-0.6$, $-0.65$ and $-0.7$), $\lambda^{\bm{Q}}$ becomes maximum at ${\bm{Q}}=(Q_{\rm pp1},\pi)$ ($(Q_{\rm pp2},0)$) as shown in Fig. \ref{Fig5} (b) ((c)) and reaches unity at $T_c$ where the second-order phase transition for the FFLO SC state with the finite COM momentum of Cooper pairs $(Q_{\rm ph1},0)$ ($(Q_{\rm ph2},\pi)$) takes place. 
Therefore the enhancement of the FFLO excitonic fluctuations help the increase of $T_c$ through the strong enhancement of pairing interaction yielding finite the FFLO Cooper-pairs. 

\begin{figure}[t]
\centering
\includegraphics[width=8.7cm]{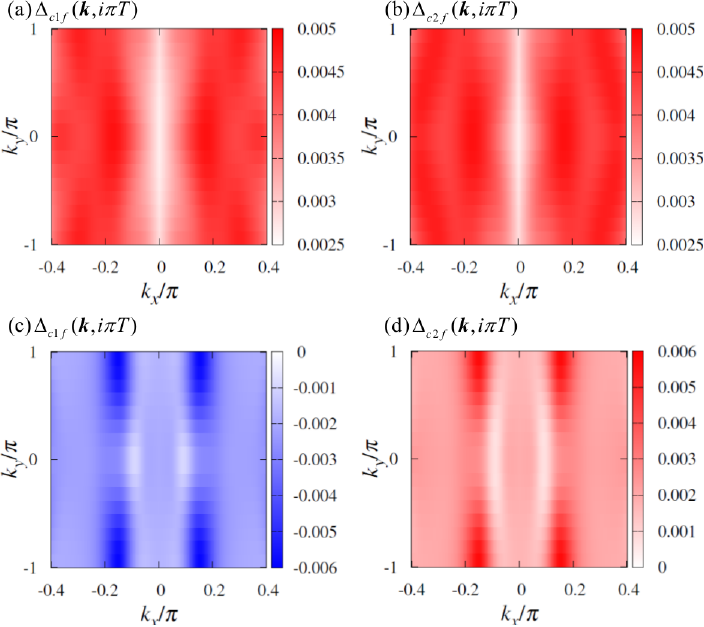}
\caption{(Color online) 
Gap functions with the lowest Matsubara frequency $\Delta^{\bm{Q}}_{\alpha f}(\bm{k},i\pi T)$ with $\alpha=c_1$ (a) and $c_2$ (b) for $D=-0.5$ at $T=0.035$, and those for $D=-0.65$ at $T=0.02$ (c) and (d). 
}
\label{Fig6}
\end{figure}

Here, we discuss the pairing mechanism of the FFLO SC state in the present system. In the case with $D=-0.5$, the bare SC susceptibility becomes maximum at ${\bm{Q}}=(Q_{\rm pp1},\pi)$ due to the particle-particle nesting between the $c_-$ and $f$ FSs  as shown in Fig. \ref{Fig4} (e). This SC fluctuation is largely enhanced by the pair scattering from $(c_-: \bm{k}\uparrow,f: -\bm{k}+\bm{Q}\downarrow)$ to  $(f: -\bm{k'}\downarrow,c_-: \bm{k'}+\bm{Q}\uparrow)$ due to the effective pairing potential $\hat{U}(k,k',\bm{Q})$ which has a large positive value for $\bm{k}-\bm{k'}\sim(Q_{\rm ph1},0)$ and divergently increases when approaching $T_{\rm EP}$, resulting in the FFLO SC state of the $(c_-,f)$ pair with the COM momentum ${\bm{Q}}=(Q_{\rm pp1},\pi)$. As for the case with $D=-0.65$, the bare SC susceptibility becomes maximum at ${\bm{Q}}=(Q_{\rm pp2},0)$ due to the particle-particle nesting between the $c_+$ and $f$ FSs (see Fig. \ref{Fig4} (f)) and is largely enhanced by the pair scattering from $(c_+: \bm{k}\uparrow,f: -\bm{k}+\bm{Q}\downarrow)$ to  $(f: -\bm{k'}\downarrow,c_+: \bm{k'}+\bm{Q}\uparrow)$ due to $\hat{U}(k,k',\bm{Q})$ which is large positive for $\bm{k}-\bm{k'}\sim(Q_{\rm ph2},\pi)$, resulting in the FFLO SC state of the $(c_+,f)$ pair with the COM momentum ${\bm{Q}}=(Q_{\rm pp2},0)$. Actually, the obtained gap functions with $(c_1,f)$ and $(c_2,f)$ pairs have the same (opposite) sign for $D=-0.5$ ($-0.65$) as shown in Figs. \ref{Fig6} (a) and (b) ((c) and (d)), where the band representation of the gap function becomes dominant for $(c_-,f)$ ($(c_+,f)$) pair with bonding (anti-bonding) $c_{-(+)}\sim (c_1 +(-) c_2)/\sqrt{2}$.

\begin{figure}[t]
\centering
\includegraphics[width=8.0cm]{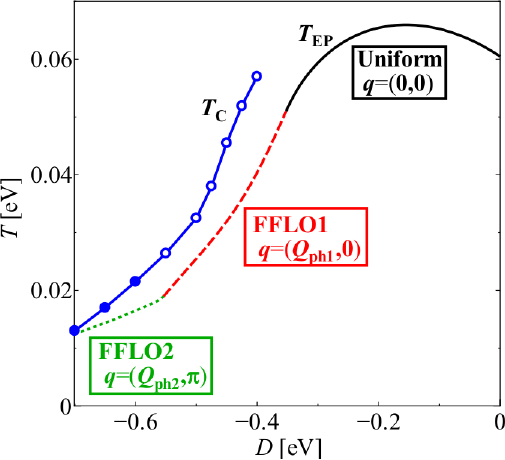}
\caption{(Color online) 
Phase diagram on the $D$-$T$ plane for $V=0.6$ and $\langle n\rangle=2$, where $T_c$ is the transition temperature for the FFLO superconductivity with ${\bm{Q}}=(Q_{\rm pp1},\pi)$ (open circles) and ${\bm{Q}}=(Q_{\rm pp2},0)$ (closed circles), and $T_{\rm EP}$ is that for the excitonic phase of the uniform (solid line), FFLO1 (dashed line) and FFLO2 (dotted line) states.
}
\label{Fig7}
\end{figure}

Finally, 
we summarize the transition temperatures of $T_{\rm EP}$ and $T_{\rm c}$ as a function of $D$ in Fig. \ref{Fig7}. 
When $T$ decreases, the excitonic susceptibilities $[\hat{\chi}^c(\bm{q},0)]_{\alpha f,\alpha f}$ with $\bm{q}=(0,0)$, $(Q_{\rm ph1},0)$ and $(Q_{\rm ph2},\pi)$ (see Fig. \ref{Fig4}) diverge at the critical temperature $T_{\rm EP}$ below which the uniform, FFLO1 and FFLO2 EPs are realized\cite{JPSJ.87.054701} for $D>-0.35$, $-0.55<D<-0.35$ and $D<-0.55$, respectively. 
When approaching $T_{\rm EP}$, the excitonic fluctuations are largely enhanced and mediate the FFLO superconductivity with ${\bm{Q}}=(Q_{\rm pp1},\pi)$ near the FFLO1 EP and that with ${\bm{Q}}=(Q_{\rm pp2},0)$ near the FFLO2 EP. 
Such the FFLO nesting mechanism can be possible for the high-pressured semimetallic situation of Ta$_2$NiSe$_5$ which is discussed in the following section. 
On the other hand in the semiconducting region with $D>-0.35$, 
the uniform excitonic fluctuation might mediate the uniform but strong-coupling pairing resulting in the BEC-type superconductivity which requires a strong-coupling theory beyond the present RPA approach. 

\section{Summary and Discussion}\label{sec4}
In summary, we have studied the quasi 1-D three-chain Hubbard model for Ta$_2$NiSe$_5$ within the RPA and have found that the FFLO superconductivity is mediated by the FFLO excitonic fluctuation which is largely enhanced towards the FFLO excitonic order realized in the semimetallic case with the imbalanced two $c$ and one $f$ FSs. 
The obtained FFLO superconductivity together with the FFLO excitonic order is expected to take place in the semimetallic Ta$_2$NiSe$_5$ under high pressure. 

In this paper, we concentrate ourselves on the excitonic fluctuations and their SC mechanism within the MF-RPA theory 
in the semimetallic regime with the high-pressure and low temperature region 
where the BCS weak coupling treatment of the exciton-pairs and Cooper-pairs is justified. 
Hence the strong coupling theory like the dynamical mean-field theory is needed for the 
semiconducting regime with the ambient (low-pressure) and high temperature region 
where actually discussed by the experiments\cite{PRB.97.115154,TrARPES} and theoretical calculations\cite{PRB.90.155116,PRL.120.247602}.
Recently, we have also studied the same model with finite $U$ and $V$ on the basis of the dynamical mean-field theory (DMFT) and have found that the excitonic fluctuation dominates over the other fluctuations due to the strong enhancement of the local excitonic fluctuation and then yields the psedogap-like behavior in the conduction band dispersion. 
Explicit results with the DMFT together with the results of the strong-coupling pairing in the BEC regime will be presented in the near future. 

Clear differences between the uniform, FFLO1 and FFLO2 excitonic phases obtained in this study are the ordered wavevectors $\bm{q}$=$(0,0)$, $(Q_{\rm ph1},0)$ and $(Q_{\rm ph2},\pi)$, respectively. 
Hence, by checking such the conduction-valence band nesting with the corresponding wavevectors, those excitonic states can be distinguished. 
Direct observation of the band dispersion by the ARPES experiment is one way to confirm the types of excitonic phases. We have already shown that the band dispersion in the FFLO phase becomes semimetallic both the FF\cite{JPSJ.85.053703} and LO states\cite{JPSJ.87.054701} in contrast to that in the uniform one which becomes fully gapped semiconducting and then we expect that the transport properties of the FFLO phases are different from those in the uniform phase. 
In addition, the non-trivial phase of the complex order parameters obtained in the FFLO states\cite{JPSJ.87.054701} is considered to be responsible for a kind of anomaly in the hole conductivity. 
Moreover, the optical response such as the resonant inelastic X-ray scattering\cite{PPB.98.035149} 
can become a good probe for detecting the enhancement of the excitonic dispersion when the exciton condensation occurs in the strong coupling regime. 
The significant softening of the particle-hole excitation with the wavevectors $\bm{q}$=$(0,0)$, $(Q_{\rm ph1},0)$ and $(Q_{\rm ph2},\pi)$ would be observed towards each instability for the uniform, FFLO1 and FFLO2 phases, respectively. 


For the realization of the present FFLO mechanism in actual system, 
it is essential that two conduction FSs exist in the high-pressure electronic states of Ta$_2$NiSe$_5$ 
as mentioned in Sec. \ref{sec3}, 
and in particular the number of $c$-band FS in the high-pressure region around $T_{C}$ is still 
unsettled. 
In addition, the impurity effect on the EPs could rapidly decreases $T_{\rm EP}$ as studied by Zittartz\cite{PR.164.575} 
where the excitonic phase is easily destroyed and $T_{\rm EP}$ could rapidly decreases by the nonmagnetic impurity as same as the magnetic impurity effect on the conventional superconductivity. 
Therefore there are two possible FFLO SC scenarios under the assumption that $T_{S}$ is $T_{\rm EP}$; 
One is the FFLO1 scenario where the disorder is strong enough to suppress $T_{\rm EP}$ with relatively small $|D|$ corresponding to the FFLO1 region, while the other is the FFLO2 scenario where the disorder is relatively weak and $T_{\rm EP}$ survives in the FFLO1 region but disappears around the large-$|D|$ FFLO2 region. 
Which scenarios realize in the real material will depend on the strength of the disorder and such the study is an important future work. 

The effect of the uniform monoclinic distortion\cite{PRB.87.035121,PRB.93.041105} 
would be still relevant for the high-pressured low-temperature phase\cite{Nakano2018,Matsubayashi}. 
We check this effect in the present theory by adding a $c$-$f$ hopping $t_{cf}$, 
and confirm that the enhancement of the FFLO excitonic fluctuation is almost unchanged but the uniform fluctuation is slightly suppressed. 
Furthermore, in our previous MF study on the same model\cite{JPSJ.87.054701}, 
the coexisted excitonic state with the uniform ($\bm{q}=\bm{0}$) and FFLO ($\bm{q}\neq\bm{0}$) wavevector is stabilized against the both single-$\bm{q}$ state of the uniform and FFLO state
which is interesting in analogy with the multi-$\bm{q}$ phase of the FFLO superconductivity\cite{JPSJ.67.736}. 

The relation between the excitonic order and the uniform distortion including the phonon coupling 
together with the formation of the antiferroelectric and the ferroelectric toroidal moments, 
which is recently observed \cite{PRB.98.045139}, 
is also important topic and should be clarified 
by the more microscopic analysis based on the effective Wannier-orbital model for the distorted low-temperature phase of Ta$_2$NiSe$_5$, and 
such the study is now under the way.



\begin{acknowledgment}
The authors give a special thanks to prof. Hidetoshi Fukuyama for providing interesting idea and variable comments and discussions. 
This research was supported in part by Multidisciplinary Cooperative Research Program in CCS, University of Tsukuba. 
This work was partially supported by a Grant-in-Aid for Scientific Research from 
the Ministry of Education, Culture, Sports, Science and Technology. 
\end{acknowledgment}


\bibliography{69156_cond}


\appendix
\section{Intersite $c$-$f$ Coulomb interaction Hamiltonian \& MF approximation}\label{sec-V}
Here we summarize the expression of the intersite $c$-$f$ Coulomb interaction in the present three-chain model which can be written by, 
\begin{align}
H_{V}
&=V\sum_{\bm{i}\alpha}\left(n^{c}_{\bm{i}\alpha}+n^{c}_{\bm{i}-\bm{a}\alpha}\right)n^{f}_{\bm{i}}\\
&=\frac{V}{N}\sum_{\{\bm{k}_i\}}\sum_{\alpha\sigma\sigma\rq{}}\phi(\bm{k}_1-\bm{k}_2)
c^{\dagger}_{\bm{k}_1\alpha\sigma}c_{\bm{k}_2\alpha\sigma}f^{\dagger}_{\bm{k}_3\sigma\rq{}}f_{\bm{k}_4\sigma\rq{}}\nonumber\\
&\quad\times\delta(\bm{k}_1-\bm{k}_2+\bm{k}_3-\bm{k}_4),
\end{align}
where $\bm{\delta}$ is the inter-unitcell vectors between Ta-Ni as $\bm{\delta}=\bm{0},-\bm{a}_1$ and thus $\phi(\bm{k})=\sum_{\bm{\delta}}e^{-i\bm{k}\cdot\bm{\delta}}=1+e^{i\bm{k}\cdot\bm{a}_1}$. 
Within the MF approximation, 
this Hamiltonian can be written with three distinct MF order parameters: 
(i) density-type, (ii) hybridization-type, and (iii) pair-formation-type as shown in the following subsections. 

\subsection{(i) Density-type MF}
The density-type MF Hamiltonian $H_{V}^{\rm den}$ and their order parameters $\Delta_{c}^{\rm den}(\bm{q})$ and $\Delta_{f}^{\rm den}(\bm{q})$ can be written by, 
\begin{align}
&H_{V}^{\rm den}\!\!=\!\!-\!\sum_{\bm{q}\bm{k}\sigma}\!
\biggl(\!
\sum_{\alpha}
\Delta_{f}^{\rm den}(\bm{q})c^{\dagger}_{\bm{k}\alpha\sigma}c_{\bm{k}+\bm{q}\alpha\sigma}
\!+\!\Delta_{c}^{\rm den}(\bm{q})f^{\dagger}_{\bm{k}\sigma}f_{\bm{k}+\bm{q}\sigma}\!
\biggr)\nonumber\\
&\qquad\quad+NV\sum_{\bm{q}}\phi(\bm{q})\langle\rho_{\bm{q}}^{c}\rangle\langle\rho_{\bm{-q}}^{f}\rangle
,\label{eq:HMF-d}\\
&\Delta_{c}^{\rm den}(\bm{q})\!=\!\!-\!\frac{V}{N}\sum_{\bm{k}\alpha\sigma}\phi(\bm{q})\langle c^{\dagger}_{\bm{k}\alpha\sigma}c_{\bm{k}+\bm{q}\alpha\sigma}\rangle\!=\!V\phi(\bm{q})\langle\rho_{\bm{q}}^{c}\rangle,\\
&\Delta_{f}^{\rm den}(\bm{q})\!=\!\!-\!\frac{V}{N}\sum_{\bm{k}\sigma}\phi(\bm{q})\langle f^{\dagger}_{\bm{k}+\bm{q}\sigma}f_{\bm{k}\sigma}\rangle\!=\!V\phi(\bm{q})\langle\rho_{\bm{q}}^{f}\rangle,\label{eq:op-d}
\end{align}
where $c$ and $f$ electron density operators are written as follows,
\begin{align}
\rho_{\bm{q}}^{c}&=\frac{1}{N}\sum_{\bm{k}\alpha\sigma}c^{\dagger}_{\bm{k}\alpha\sigma}c_{\bm{k}+\bm{q}\alpha\sigma},~~
\rho_{\bm{q}}^{f}=\frac{1}{N}\sum_{\bm{k}\sigma}f^{\dagger}_{\bm{k}\sigma}f_{\bm{k}+\bm{q}\sigma}, 
\end{align}
where their MF expectations $\langle\rho_{\bm{q}}^{c}\rangle$,  $\langle\rho_{\bm{q}}^{f}\rangle$ correspond to the order parameters for the $c$- and $f$-electron charge density wave (CDW) with their modulated wavevector $\bm{q}\neq\bm{0}$
or for the phase separation $\bm{q}=\bm{0}$. 
In normal phase, they become zero $\langle\rho_{\bm{q}\neq\bm{0}}^{c(f)}\rangle=0$
except for $\bm{q}=\bm{0}$ corresponding to the total $c(f)$ electron number per unitcell $\langle\rho_{\bm{q}=\bm{0}}^{c(f)}\rangle=\langle n^{c(f)}\rangle$. 
As seen in the last term of Eq.(\ref{eq:HMF-d}), 
$\Delta_{c}^{\rm den}(\bm{q})$ and $\Delta_{f}^{\rm den}(\bm{q})$ become finite at the same time 
and thus the $c$-CDW and $f$-CDW orders occur simultaneously due to the $c$-$f$ intersite Coulomb interaction $V$.

\subsection{(ii) Hybridization-type MF}
Similarly, the hybridization-type MF Hamiltonian $H_{V}^{\rm hyb}$ and the order parameters $\Delta_{\alpha\sigma\sigma\rq{}}^{\rm hyb}(\bm{k},\bm{q})$ are given by, 
\begin{align}
&H_{V}^{\rm hyb}\!\!=\!\!
\sum_{\bm{qk}\alpha}\sum_{\sigma\sigma\rq{}}
\biggl[
\Delta_{\alpha\sigma\sigma\rq{}}^{\rm hyb}(\bm{k},\bm{q})c^{\dagger}_{\bm{k}\alpha\sigma}f_{\bm{k}+\bm{q}\sigma\rq{}}
+h.c.\nonumber\\
&\qquad-V\phi(\bm{k}\rq{}-\bm{k})
\langle c^{\dagger}_{\bm{k}\rq{}\alpha\sigma}f_{\bm{k}\rq{}+\bm{q}\sigma\rq{}}\rangle
\langle f^{\dagger}_{\bm{k}+\bm{q}\sigma\rq{}}c_{\bm{k}\alpha\sigma}\rangle
\biggr]
,\label{eq:HMF-h}\\
&\Delta_{\alpha\sigma\sigma\rq{}}^{\rm hyb}(\bm{k},\bm{q})=-\frac{V}{N}\sum_{\bm{k}\rq{}}\phi(\bm{k}-\bm{k}\rq{})\langle f^{\dagger}_{\bm{k}\rq{}+\bm{q}\sigma\rq{}}c_{\bm{k}\rq{}\alpha\sigma}\rangle,\\
&\Delta_{\alpha\sigma\sigma\rq{}}^{\rm hyb*}(\bm{k},\bm{q})=-\frac{V}{N}\sum_{\bm{k}\rq{}}\phi(\bm{k}\rq{}-\bm{k})\langle c^{\dagger}_{\bm{k}\rq{}\alpha\sigma}f_{\bm{k}\rq{}+\bm{q}\sigma\rq{}}\rangle,\label{eq:op-h}
\end{align}
where the finite expectation values of $\langle f^{\dagger}_{\bm{k}+\bm{q}\sigma\rq{}}c_{\bm{k}\alpha\sigma}\rangle$, $\langle c^{\dagger}_{\bm{k}\alpha\sigma}f_{\bm{k}+\bm{q}\sigma\rq{}}\rangle$ 
indicate the intersite $c$-$f$ bond formation 
and also correspond to a spontaneous phase transition towards the EP 
with the exciton-pairs COM wavevector $\bm{q}$ since $H_{0}$ has no $c$-$f$ direct hopping. 

\subsection{(iii) Pair-formation-type MF}
Finally, the pair-formation-type MF Hamiltonian $H_{V}^{\rm pair}$ and the order parameters $\Delta_{\alpha\sigma\sigma\rq{}}^{\rm pair}(\bm{k},\!\bm{q})$ are given by, 
\begin{align}
&H_{V}^{\rm pair}\!\!=\!\!-\sum_{\bm{qk}\alpha}\sum_{\sigma\sigma\rq{}}
\biggl[
\Delta_{\alpha\sigma\sigma\rq{}}^{\rm pair}(\bm{k},\bm{q})c^{\dagger}_{\bm{k}\alpha\sigma}f^{\dagger}_{-\bm{k}+\bm{q}\sigma\rq{}}
+h.c.\nonumber\\
&\quad-V\phi(\bm{k}\rq{}\!-\!\bm{k})
\langle c^{\dagger}_{\bm{k}\rq{}\alpha\sigma}f^{\dagger}_{-\bm{k}\rq{}+\bm{q}\sigma\rq{}}\rangle
\langle f_{-\bm{k}+\bm{q}\sigma\rq{}}c_{\bm{k}\alpha\sigma}\rangle
\biggr]
,\label{eq:HMF-p}\\
&\Delta_{\alpha}^{\rm pair}(\bm{k},\bm{q})=-\frac{V}{N}\sum_{\bm{k}\rq{}}\phi(\bm{k}-\bm{k}\rq{})\langle f_{-\bm{k}\rq{}+\bm{q}\sigma}c_{\bm{k}\rq{}\alpha\sigma}\rangle,\\
&\Delta_{\alpha}^{\rm pair*}(\bm{k},\bm{q})=-\frac{V}{N}\sum_{\bm{k}\rq{}}\phi(\bm{k}\rq{}-\bm{k})\langle c^{\dagger}_{\bm{k}\rq{}\alpha\sigma}f^{\dagger}_{-\bm{k}\rq{}+\bm{q}\sigma}\rangle,\label{eq:op-p}
\end{align}
where the expectation values of $\langle f_{-\bm{k}+\bm{q}\sigma\rq{}}c_{\bm{k}\alpha\sigma}\rangle$ and $\langle c^{\dagger}_{\bm{k}\alpha\sigma}f^{\dagger}_{-\bm{k}+\bm{q}\sigma\rq{}}\rangle$ become finite when SC state with 
the interband $c$-$f$ Cooper-pairs formation with the COM wavenumber $\bm{q}$ emerges. 

For the semiconducting and semimetallic systems, $f$-electron number is almost fully occupied, $\langle n^{f}\rangle\sim 2$, 
while $c$-electron number is nearly empty, $\langle n^{c}\rangle\sim0$, 
where $c$- and $f$-electron charge susceptibilities $\chi_{cccc}^{0}(q)$ and $\chi_{ffff}^{0}(q)$ is small in comparison with the excitonic susceptibilities $\chi_{cfcf}^{0}(q)$ and $\chi_{fcfc}^{0}(q)$ in Eq.(\ref{eq:chi0ex}). 
Hence we can neglect charge density type order for describing the weak coupling (BCS) semimetallic regime as discussed in the main text.

\section{Explicit form of extended basis function}\label{sec-ebf}
The explicit form of the extended basis function $g_{\ell\ell\rq{}}^{\alpha}(\bm{k})$ is written as follows, 
\begin{align}
g_{\ell\ell\rq{}}^{\alpha}(\bm{k})
&=a_{\ell\ell\rq{}}^{\alpha(0)}+a_{\ell\ell\rq{}}^{\alpha(+)}e^{i\bm{k}\cdot\bm{a}_1}+a_{\ell\ell\rq{}}^{\alpha(-)}e^{-i\bm{k}\cdot\bm{a}_1},\\
a_{\ell\ell\rq{}}^{\alpha(0)}
&=
\!\delta_{\alpha 1}\delta_{\ell1}\delta_{\ell\rq{}1}
\!+\delta_{\alpha 2}\delta_{\ell1}\delta_{\ell\rq{}2}
\!+\delta_{\alpha 3}\delta_{\ell1}\delta_{\ell\rq{}3}\nonumber\\
&+\delta_{\alpha 4}\delta_{\ell2}\delta_{\ell\rq{}1}
\!+\delta_{\alpha 5}\delta_{\ell2}\delta_{\ell\rq{}2}
\!+\delta_{\alpha 6}\delta_{\ell2}\delta_{\ell\rq{}3}\nonumber\\
&+\delta_{\alpha 7}\delta_{\ell3}\delta_{\ell\rq{}1}
\!+\delta_{\alpha 8}\delta_{\ell3}\delta_{\ell\rq{}2}
\!+\delta_{\alpha 9}\delta_{\ell3}\delta_{\ell\rq{}3},\\
a_{\ell\ell\rq{}}^{\alpha(+)}
&=
\delta_{\alpha 10}\delta_{\ell1}\delta_{\ell\rq{}3}
\!+\delta_{\alpha 11}\delta_{\ell2}\delta_{\ell\rq{}3},\\
a_{\ell\ell\rq{}}^{\alpha(-)}
&=
\!\delta_{\alpha 12}\delta_{\ell3}\delta_{\ell\rq{}1}
\!+\delta_{\alpha 13}\delta_{\ell3}\delta_{\ell\rq{}2},
\end{align}
where $\alpha$=1-9 corresponds to the intra-cell orbital sets $(\ell,\ell\rq{})$ 
and $\alpha$=10-13 the inter-cell $c$-$f$ orbital sets as
$10:(\ell,\ell\rq{})$=$(c_1,f)$, 
$11:(\ell,\ell\rq{})$=$(c_2,f)$, 
$12:(\ell,\ell\rq{})$=$(f,c_1)$, 
$13:(\ell,\ell\rq{})$=$(f,c_2)$. 
\section{MF Phase boundary condition}\label{sec-pbc}
In this Appendix, we give the formulation of the excitonic and superconducting phase boundary within the MF approximation for the intersite $c$-$f$ Coulomb interaction $V$ in the quasi three-chain Hubbard model 
and we also shows the superconducting phase boundary in the $D$-$T$ plane 
which can be obtained for the negative $c$-$f$ Coulomb $V<0$ which corresponds to the attractive pairing interaction between $c$ and $f$ electrons effectively. 

\subsection{Bare Green\rq{}s functions \& bare susceptibilities for the three-chain model}
The bare Green\rq{}s functions for the present quasi 1-D three-chain model can be written as follows,
\begin{align}
&G^{0}_{c_1c_1}(\bm{k},i\varepsilon_n)=G^{0}_{c_2c_2}(\bm{k},i\varepsilon_n)\nonumber\\
&\qquad\qquad\quad~=\frac{1}{2}\left(\frac{1}{i\varepsilon_n-\zeta_{\bm{k}+}^{c}}+\frac{1}{i\varepsilon_n-\zeta_{\bm{k}-}^{c}}\right),\\
&G^{0}_{ff}(\bm{k},i\varepsilon_n)=\frac{1}{i\varepsilon_n-\zeta_{\bm{k}}^{f}},\\
&G^{0}_{c_1c_2}(\bm{k},i\varepsilon_n)=\frac{\eta_{\bm{k}}^{c}}{2}\left(\frac{1}{i\varepsilon_n-\zeta_{\bm{k}+}^{c}}-\frac{1}{i\varepsilon_n-\zeta_{\bm{k}-}^{c}}\right),
\end{align}
where $\zeta_{\bm{k}}^{f}=\varepsilon_{\bm{k}f}-\mu$, $\zeta_{\bm{k}\pm}^{c}=\varepsilon_{\bm{k}c\pm}-\mu$, and
$\eta_{\bm{k}}^{c}=\varepsilon_{\bm{k}c}\rq{}/|\varepsilon_{\bm{k}c}\rq{}|$. 
The bare susceptibilities for particle-hole and particle-particle channels $\chi^{0}$ and 
$\psi^{0}$ are also written as follows,
\begin{align}
&\chi_{\ell\ell\rq{}mm\rq{}}^{0}(\bm{q},i\omega_\nu) \nonumber\\
&=-\frac{T}{N}\sum_{\bm{k}n}G_{m\ell}^{0}(\bm{k},i\varepsilon_n)G_{\ell\rq{}m\rq{}}^{0}(\bm{k}+\bm{q},i\varepsilon_n+i\omega_\nu),\label{eq:chi0ex}\\
&\psi_{\ell\ell\rq{}mm\rq{}}^{0}(\bm{q},i\omega_\nu) \nonumber\\
&=\frac{T}{N}\sum_{\bm{k}n}G_{\ell m}^{0}(\bm{k},i\varepsilon_n)G_{\ell\rq{}m\rq{}}^{0}(-\bm{k}+\bm{q},-i\varepsilon_n+i\omega_\nu),\label{eq:psi0sc}
\end{align}
where $\ell$ is the site-orbital index $\ell=c_1,c_2,f$. 
Especially for the semiconducting and semimetallic situation like the present system, the $c$-$f$ coupled terms with $i\omega_\nu=0$ are exclusively important for the excitonic fluctuation, 
\begin{align}
&\chi_{\alpha f\alpha f}^{0}(\bm{q},0)
=\frac{1}{2N}\sum_{\bm{k}}
\biggl(
\frac{n_{\rm F}(\zeta_{\bm{k}+\bm{q}}^{f})-n_{\rm F}(\zeta_{\bm{k}+}^{c})}{\zeta_{\bm{k}+}^{c}-\zeta_{\bm{k}+\bm{q}}^{f}}
\nonumber\\
&\qquad\qquad\qquad\qquad
+
\frac{n_{\rm F}(\zeta_{\bm{k}+\bm{q}}^{f})-n_{\rm F}(\zeta_{\bm{k}-}^{c})}{\zeta_{\bm{k}-}^{c}-\zeta_{\bm{k}+\bm{q}}^{f}}
\!\biggr)
,\\
&\psi_{\alpha f\alpha f}^{0}(\bm{q},0)
=\frac{1}{2N}\sum_{\bm{k}}
\biggl(\!
\frac{n_{\rm F}(-\zeta_{-\bm{k}+\bm{q}}^{f})-n_{\rm F}(\zeta_{\bm{k}+}^{c})}{\zeta_{\bm{k}+}^{c}+\zeta_{-\bm{k}+\bm{q}}^{f}}
\nonumber\\
&\qquad\qquad\qquad\qquad
+
\frac{n_{\rm F}(-\zeta_{-\bm{k}+\bm{q}}^{f})-n_{\rm F}(\zeta_{\bm{k}-}^{c})}{\zeta_{\bm{k}-}^{c}+\zeta_{-\bm{k}+\bm{q}}^{f}}
\!\biggr),
\end{align}
where $n_{\rm F}(x)$ is Fermi distribution function $n_{\rm F}(x)=1/(e^{\beta x}+1)$. 
Both susceptibilities become quite large and/or diverge at low $T$ when the $c$-$f$ nesting conditions are satisfied in the semimetallic situation. 

\subsection{Excitonic phase boundary}
Here we derive the EP boundary condition within the MF approximation. For simplicity, we drop the spin-dependence of the order parameter in Eq.(\ref{eq:op-h}) as $\Delta_{\alpha\sigma\sigma\rq{}}^{\rm hyb}(\bm{k},\bm{q})=\delta_{\sigma\sigma\rq{}}\Delta_{\alpha}^{\rm ex}(\bm{k},\bm{q})$ since there is no magnetic sign in the present semimetallic system. 
Hence the excitonic order parameter $\Delta_{\alpha}^{\rm ex}(\bm{k},\bm{q})$ can be written as follows, 
\begin{align}
\Delta_{\alpha}^{\rm ex}(\bm{k},\bm{q})
&=-\frac{V}{N}\sum_{\bm{k}\rq{}}\phi(\bm{k}-\bm{k}\rq{})G_{\alpha f}^{~\bm{q}}(\bm{k}\rq{},\tau=-0_{+})
,\label{eq:op-h1}
\end{align}
where $G_{\alpha f}^{~\bm{q}}(\bm{k},\tau)$ is the $c$-$f$ anomalous Green\rq{}s function and becomes finite (zero) in excitonic (normal) phase which can be written by the $\varepsilon_n$-representation as follows, 
\begin{align}
G_{\alpha f}^{~\bm{q}}(\bm{k},i\varepsilon_n)&=-\int_0^{\beta}d\tau \langle \mathcal{T}c_{\bm{k}\alpha\sigma}(\tau)f^{\dagger}_{\bm{k}+\bm{q}\sigma}(0)\rangle e^{i\varepsilon_n\tau}.
\label{eq:Fourier-w}
\end{align}

Around the EP boundary, $G_{\alpha f}^{~\bm{q}}(\bm{k},i\varepsilon_n)$ and $\Delta_{\alpha}^{\rm ex}(\bm{k},\bm{q})$ in Eq.(\ref{eq:op-h1}) are self-consistently determined so as to satisfy the linearized Dyson-Gorkov equations 
given by,
\begin{align}
&G_{\alpha f}^{~\bm{q}}(\bm{k},i\varepsilon_n)=G_{\alpha f}^{0\bm{q}}(\bm{k},i\varepsilon_n)\nonumber\\
&\quad+
\sum_{\alpha\rq{}}G_{\alpha\alpha\rq{}}^{0}(\bm{k},i\varepsilon_n)\Delta_{\alpha\rq{}}^{\rm ex}(\bm{k},\bm{q})G_{ff}^{0}(\bm{k}+\bm{q},i\varepsilon_{n}),\label{eq:LDG-h}\\
&\Delta_{\alpha}^{\rm ex}(\bm{k},\bm{q})=-\frac{VT}{N}\sum_{\bm{k}\rq{}n}\sum_{\alpha\rq{}}\phi(\bm{k}-\bm{k}\rq{})G_{\alpha\alpha\rq{}}^{0}(\bm{k}\rq{},i\varepsilon_n)\nonumber\\
&\qquad\qquad\qquad\times\Delta_{\alpha\rq{}}^{\rm ex}(\bm{k}\rq{},\bm{q})G_{ff}^{0}(\bm{k}\rq{}+\bm{q},i\varepsilon_n),\label{eq:sce-h}
\end{align}
where in the first line we use $G_{\alpha f}^{0\bm{q}}(\bm{k},i\varepsilon_n)=0$ in the normal phase. 
In Eq.(\ref{eq:sce-h}), the $\bm{k}$-dependence of $\Delta_{\alpha}^{\rm ex}(\bm{k},\bm{q})$ is given by $\phi(\bm{k}-\bm{k}\rq{})$ and the frequency dependence of $\Delta_{\alpha}^{\rm ex}(\bm{k},\bm{q})$ is dropped 
within the MF approximation for the static intersite Coulomb interaction $V$. 
Thus we can set the following form in general, 
\begin{align}
\Delta_{\alpha}^{\rm ex}(\bm{k},\bm{q})=\Delta_{\bm{q}\alpha}^{{\rm ex}(0)}+e^{i\bm{k}\cdot\bm{a}_1}\Delta_{\bm{q}\alpha}^{{\rm ex}(1)},\label{eq:op-form}
\end{align}
where $\Delta_{\bm{q}\alpha}^{{\rm ex}(0)}(\Delta_{\bm{q}\alpha}^{{\rm ex}(1)})$ is the complex order parameter between the $f$-site and $c$-site in the same (left neighboring) unitcell. 
By substituting Eq.(\ref{eq:op-form}) into Eq.(\ref{eq:sce-h}) 
we can obtain the simultaneous linear equations for four order parameters, 
\begin{align}
&V\left(
\begin{array}{cccc}
\chi_{0}(\bm{q})    & \chi_{1}(\bm{q})     & \chi_{2}(\bm{q})   & \chi_{4}(\bm{q}) \\
\chi_{1}^{*}(\bm{q}) & \chi_{0}(\bm{q})    & \chi_{3}(\bm{q})    & \chi_{2}(\bm{q}) \\
\chi_{2}^{*}(\bm{q}) & \chi_{3}^{*}(\bm{q}) & \chi_{0}(\bm{q})  & \chi_{1}(\bm{q}) \\
\chi_{4}^{*}(\bm{q}) & \chi_{2}^{*}(\bm{q}) & \chi_{1}^{*}(\bm{q}) & \chi_{0}(\bm{q}) \\
\end{array}
\right)
=
\left(
\begin{array}{cccc}
1&0&0&0\\
0&1&0&0\\
0&0&1&0\\
0&0&0&1\\
\end{array}
\right)\nonumber\\
&\Longrightarrow 
V\hat{M}_{\rm ex}(\bm{q})=\hat{1},\label{eq:mtx-h1}
\end{align}
where $\chi_{i}(\bm{q})(i=0\sim 4)$ is the bare particle-hole susceptibilities like Eq.(\ref{eq:chi0ex}) which are explicitly given by,
\begin{align}
&\chi_0(\bm{q})=\frac{1}{N}\sum_{\bm{k}}\chi^{0}_{\alpha f\alpha f}(\bm{k},\bm{q}),\\
&\chi_1(\bm{q})=\frac{1}{N}\sum_{\bm{k}}e^{i\bm{k}\cdot\bm{a}_1}\chi^{0}_{\alpha f\alpha f}(\bm{k},\bm{q}),\\
&\chi_2(\bm{q})=\frac{1}{N}\sum_{\bm{k}}\chi^{0}_{c_2fc_1f}(\bm{k},\bm{q}),\\
&\chi_3(\bm{q})=\frac{1}{N}\sum_{\bm{k}}e^{-i\bm{k}\cdot\bm{a}_1}\chi^{0}_{c_2fc_1f}(\bm{k},\bm{q}),\\
&\chi_4(\bm{q})=\frac{1}{N}\sum_{\bm{k}}e^{i\bm{k}\cdot\bm{a}_1}\chi^{0}_{c_2fc_1}(\bm{k},\bm{q}),
\end{align}
where $\chi^{0}_{\alpha f\alpha\rq{}f}(\bm{k},\bm{q})$=$-T\sum_{n}G^{0}_{\alpha\rq{}\alpha}(\bm{k},i\varepsilon_n)G^{0}_{ff}(\bm{k}+\bm{q},i\varepsilon_n)$. 

Then we can obtain four eigenvalues of the particle-hole susceptibility matrix $\hat{M}_{\rm ex}(\bm{q})$ 
analytically by using Ferrari's method for the quartic equation for any given $T$, $D$ and $\bm{q}$. 
The maximum eigenvalue of $\hat{M}_{\rm ex}(\bm{q})$ corresponds to the bare excitonic susceptibility $\chi_{\rm ex}^{0}(\bm{q})$
which yields the EP boundary condition $V\chi_{\rm ex}^{0}(\bm{q})=1$. 
If we drop $\chi_{i}(\bm{q})(i=2\sim 4)$ in Eq.(\ref{eq:mtx-h1}), 
the condition for the previous purely 1-D model\cite{JPSJ.85.053703,JPSJ.85.065005}
returns as $V\left(\chi_{0}(\bm{q})+|\chi_{1}(\bm{q})|\right)=1$.

\subsection{SC phase boundary}
Next we derive the SC phase boundary condition within the MF approximation in similar way of the EP case. 
Here we consider only the spin-singlet gap function in Eq.(\ref{eq:op-p}) as $\Delta_{\alpha\sigma\sigma\rq{}}^{\rm pair}(\bm{k},\bm{q})=(1-\delta_{\sigma\sigma\rq{}})\Delta_{\alpha}^{\rm sc}(\bm{k},\bm{q})$ which is written by, 
\begin{align}
\Delta_{\alpha}^{\rm sc}(\bm{k},\bm{q})&
=-\frac{V}{N}\sum_{\bm{k}\rq{}}\phi(\bm{k}-\bm{k}\rq{})F_{\alpha f}^{~\bm{q}}(\bm{k}\rq{},\tau=-0_{+})
,\label{eq:op-p1}
\end{align}
where $F_{\alpha f}^{~\bm{q}}(\bm{k},\tau)$ is the $c$-$f$ anomalous Green\rq{}s function for SC phase 
and  becomes finite (zero) in superconducting (normal) phase which is given by the $\varepsilon_n$-representation as follows, 
\begin{align}
F_{\alpha f}^{~\bm{q}}(\bm{k},i\varepsilon_n)=-\int_0^{\beta}\!d\tau \langle \mathcal{T}c_{\bm{k}\alpha\sigma}^{\dagger}(\tau)f^{\dagger}_{\bm{k}+\bm{q}\sigma}(0)\rangle e^{i\varepsilon_n\tau}.
\end{align}

Around the SC phase boundary, $F_{\alpha f}^{~\bm{q}}(\bm{k},i\varepsilon_n)$ and $\Delta_{\alpha}^{\rm sc}(\bm{k},\bm{q})$ in Eq.(\ref{eq:op-p1}) are self-consistently determined so as to satisfy the linearized Dyson-Gorkov equations which are 
given by,
\begin{align}
&F_{\alpha f}^{~\bm{q}}(\bm{k},i\varepsilon_n)\!=F_{\alpha f}^{0\bm{q}}(\bm{k},i\varepsilon_n)\nonumber\\
&~+\sum_{\alpha\rq{}}\!G_{\alpha\alpha\rq{}}^{0}(\bm{k},i\varepsilon_n)\Delta_{\alpha\rq{}}^{\rm sc}(\bm{k},\bm{q})G_{ff}^{0}(-\bm{k}\!+\!\bm{q},-i\varepsilon_{n}),\label{eq:LDG-p}\\
&\Delta_{\alpha}^{\rm sc}(\bm{k},\bm{q})=-\frac{VT}{N}\sum_{\bm{k}\rq{}n}\sum_{\alpha\rq{}}\phi(\bm{k}-\bm{k}\rq{})G_{\alpha\alpha\rq{}}^{0}(\bm{k}\rq{},i\varepsilon_n)\nonumber\\
&\qquad\qquad\qquad\times\Delta_{\alpha\rq{}}^{\rm sc}(\bm{k}\rq{},\bm{q})G_{ff}^{0}(-\bm{k}\rq{}\!+\!\bm{q},-i\varepsilon_n)
,\label{eq:sce-p}
\end{align}
where in the first line we use $F_{\alpha f}^{0\bm{q}}(\bm{k},i\varepsilon_n)=0$ in the normal phase. 
In Eq.(\ref{eq:sce-p}), the $\bm{k}$-dependence of $\Delta_{\alpha}^{\rm sc}(\bm{k},\bm{q})$ is given by $\phi(\bm{k}-\bm{k}\rq{})$ and the frequency dependence of $\Delta_{\alpha}^{\rm sc}(\bm{k},\bm{q})$ is dropped 
for the same reason of the previous subsection. 
The explicit form of $\Delta_{\alpha}^{\rm sc}(\bm{k},\bm{q})$ is applicable as the same form of $\Delta_{\alpha}^{\rm ex}(\bm{k},\bm{q})$, 
\begin{align}
\Delta_{\alpha}^{\rm sc}(\bm{k},\bm{q})=\Delta_{\bm{q}\alpha}^{{\rm sc}(0)}+e^{i\bm{k}\cdot\bm{a}_1}\Delta_{\bm{q}\alpha}^{{\rm sc}(1)},\label{eq:op-form-p}
\end{align}
and thus we can obtain the similar equations with the EP case which is given by, 
\begin{align}
&-V\left(
\begin{array}{cccc}
\psi_{0}(\bm{q})    & \psi_{1}(\bm{q})     & \psi_{2}(\bm{q})   & \psi_{4}(\bm{q}) \\
\psi_{1}^{*}(\bm{q}) & \psi_{0}(\bm{q})    & \psi_{3}(\bm{q})    & \psi_{2}(\bm{q}) \\
\psi_{2}^{*}(\bm{q}) & \psi_{3}^{*}(\bm{q}) & \psi_{0}(\bm{q})  & \psi_{1}(\bm{q}) \\
\psi_{4}^{*}(\bm{q}) & \psi_{2}^{*}(\bm{q}) & \psi_{1}^{*}(\bm{q}) & \psi_{0}(\bm{q}) \\
\end{array}
\right)
=
\left(
\begin{array}{cccc}
1&0&0&0\\
0&1&0&0\\
0&0&1&0\\
0&0&0&1\\
\end{array}
\right)\nonumber\\
&\Longrightarrow 
-V\hat{M}_{\rm sc}(\bm{q})=\hat{1},\label{eq:mtx-p1}
\end{align}
where $\psi_{i}(\bm{q})(i=0\sim 4)$ is the bare particle-particle susceptibilities like Eq.(\ref{eq:psi0sc}) which are explicitly given by,
\begin{align}
&\psi_0(\bm{q})=\frac{T}{N}\sum_{\bm{k}}\psi^{0}_{\alpha f\alpha f}(\bm{k},\bm{q}),\\
&\psi_1(\bm{q})=\frac{T}{N}\sum_{\bm{k}}e^{i\bm{k}\cdot\bm{a}_1}\psi^{0}_{\alpha f\alpha f}(\bm{k},\bm{q}),\\
&\psi_2(\bm{q})=\frac{T}{N}\sum_{\bm{k}}\psi^{0}_{c_2fc_1f}(\bm{k},\bm{q}),\\
&\psi_3(\bm{q})=\frac{T}{N}\sum_{\bm{k}}e^{-i\bm{k}\cdot\bm{a}_1}\psi^{0}_{c_2fc_1f}(\bm{k},\bm{q}),\\
&\psi_4(\bm{q})=\frac{T}{N}\sum_{\bm{k}}e^{i\bm{k}\cdot\bm{a}_1}\psi^{0}_{c_2fc_1f}(\bm{k},\bm{q}),
\end{align}
where $\psi^{0}_{\alpha f\alpha\rq{}f}(\bm{k},\bm{q})$=$T\sum_{n}G^{0}_{\alpha\alpha\rq{}}(\bm{k},i\varepsilon_n)G^{0}_{ff}(-\bm{k}+\bm{q},i\varepsilon_n)$. 

Then we can obtain four eigenvalues of the particle-particle susceptibility matrix $\hat{M}_{\rm sc}(\bm{q})$ 
analytically as the same as the EP case for any given $T$, $D$ and $\bm{q}$. 
The maximum eigenvalue of $\hat{M}_{\rm sc}(\bm{q})$ corresponds to the bare SC susceptibility $\psi_{\rm sc}^{0}(\bm{q})$
which yields the SC boundary condition $-V\psi_{\rm sc}^{0}(\bm{q})=1$, 
where the negative $c$-$f$ interaction $V<0$ is needed because $\psi_{\rm sc}^{0}(\bm{q})>0$.

\section{SC phase boundary in $D$-$T$ plane}\label{sec-SCphase}
\begin{figure}[t]
\centering
\includegraphics[width=8.7cm]{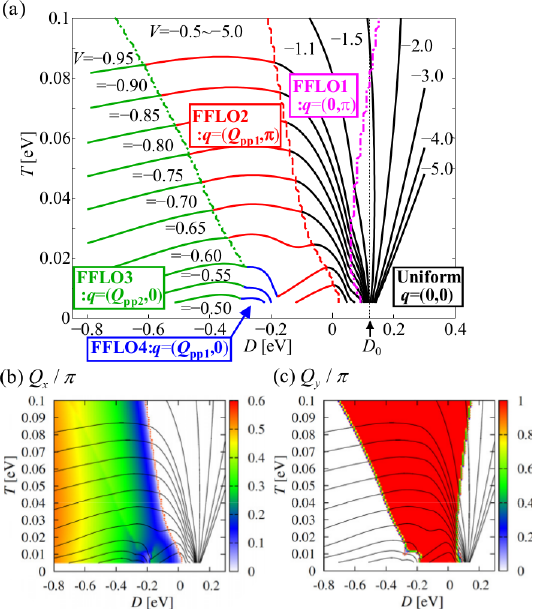}
\caption{(Color online) 
(a) $D$-$T$ phase boundaries between normal-SC phases for several values of $V<0$. 
(b),(c) The color maps for the values of $Q_x/\pi$ (b) and $Q_y/\pi$ (c) on the $D$-$T$ phase diagram. 
}
\label{Fig8}
\end{figure}
In Figure \ref{Fig8} (a), we show the SC phase boundaries with several values of $V<0$ 
in $D$-$T$ plane as well as the EP case mentioned in the main text. 
Contrary to the EP case, five distinct SC phases are observed 
; a conventional uniform SC state with $\bm{Q}=\bm{0}$, and four FFLO SC phases, FFLO1, FFLO2, FFLO3 and FFLO4 with finite $\bm{Q}=(Q_x,Q_y)$ as plotted in color maps of $Q_x$[Fig. \ref{Fig8} (b)] and $Q_y$[Fig. \ref{Fig8} (c)] as the same as the case of EP [Fig. \ref{Fig3} (b) and (c)]. 

A more complicated ($D,V$)-dependence of $T_{\rm SC}$ is observed. 
$T_{\rm SC}$ emerges around the semiconducting-semimetalic boundary $D\sim D_0$ with any strength of $|V|$, 
but its development direction is highly dependent on $|V|$; 
when $|V|$ is relatively small (weak and intermediate $|V|$-regime), $T_{\rm SC}$ increases for $D<D_0$  
while when $|V|$ is large (strong $|V|$-regime), $T_{\rm SC}$ develops for $D>D_0$. 

For weak $|V|$-regime $(|V|\siml 0.6)$, two domes of $T_{\rm SC}$ emerge as a function of $D$ 
around the shallow $|D|$ from $-0.2\siml D\siml D_0$ and the relatively deep $|D|$ from $-0.8\siml D\siml -0.2$. 
On the other hands, when $|V|$ reaches the intermediate $|V|$-regime ($|V|\simg 0.6$) 
the two domes combine into one and $T_{\rm SC}$ increases with increasing $|V|$ with exhibiting a weak $D$-dependence for wide semimetallic region. 
Such the behavior can be understood by the BCS-formula of the transition temperature, $T_{\rm SC}=\hbar\omega_{c}exp(-1/\rho_0|V|)$,  where $\omega_c$ is the energy cut-off and $|V|$ is the magnitude of the attraction and $\rho_0$ is the density of states on the FSs 
where $\rho_0$ is almost constant for $D$ in the semimetallic situation. 

In such the weak and intermediate $|V|$-regime with $D<D_0$, 
the FFLO SC states with the COM wavevectors $\bm{Q}\neq\bm{0}$ can be possible as well as the FFLO EP case 
but their wavevector $\bm{Q}$ differs from the exciton COM wavevector $\bm{q}$ in the main text, 
where the Cooper-pairs here consist of $c$-band electron with $\bm{k}$ and $f$-band electron with $-\bm{k}+\bm{Q}$
in contrast to the exciton-pairs between the $c$-FS with $\bm{k}$ and $f$-FS with $\bm{k}+\bm{q}$. 

Firstly the FFLO1 SC state is realized for the shallow $D$ region 
where $\psi^{0}_{\rm SC}(\bm{Q})$ with $\bm{Q}=(0,\pi)$ becomes dominant due to the $c_{-}$-$f$ band nesting effect. 
When the $c_{+}$-FS emerges, 
the $c$-$f$ density imbalance 
shrinks the $c_{-}$-FS by sliding in $k_x$-direction (not shown) 
so as to satisfy the particle-hole FS compensation 
which induces the $c_{-}$-$f$ band nesting with finite $Q_x$ corresponding to the FFLO2 SC states with $\bm{Q}=(Q_{\rm pp1},\pi)$ as shown in Fig. \ref{Fig8} (a) and (b). 
The further decrease of $D$ gives rise to the change of the dominant $\psi^{0}_{\rm SC}(\bm{Q})$ 
from the $c_{-}$-$f$ nesting with $\bm{Q}=(Q_{\rm pp1},\pi)$
to the $c_{+}$-$f$ nesting $\bm{Q}=(Q_{\rm pp2},0)$ 
corresponding to the emergence of the FFLO3 SC as shown in Fig. \ref{Fig8} (a) and (b). 
A fourth fluctuation FFLO4 SC with $\bm{Q}=(Q_{\rm pp1},0)$ is also observed in lower $T$ and small $|V|$ 
between the FFLO2 and FFLO3 SC  fluctuations. 

Furthermore for strong $|V|$-regime $(|V|\simg 1.5)$, 
the semiconducting-SC transition for $D>D_0$ can occur only at high temperature as shown in Fig. \ref{Fig8} (a).
In such the situation, $\psi^{0}_{\rm sc}(\bm{Q})$ become finite due to the thermal excitation 
and takes the maximum with $\bm{Q}=\bm{0}$ corresponding to the instability towards the uniform SC states.
However, $\psi^{0}_{\rm sc}(\bm{0})$ is completely suppressed with decreasing $T$ and also becomes absolutely zero at $T\rightarrow 0$ in the absence of FSs. 


\section{$\bm{q}$-dependence of bare susceptibilities of $\chi_{\rm ex}^{0}(\bm{q})$ and $\psi^{0}_{\rm sc}(\bm{q})$}\label{sec-qdep}
\begin{figure}[t]
\centering
\includegraphics[width=8.7cm]{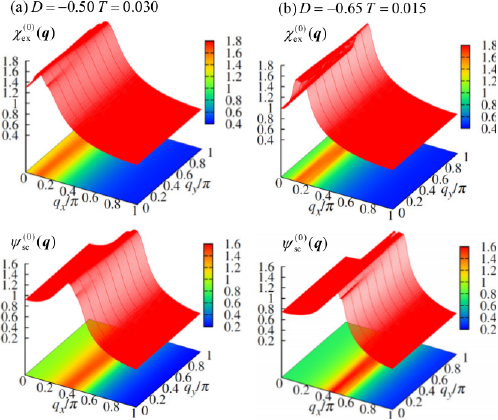}
\caption{(Color online) 
The bare excitonic susceptibility $\chi_{\rm ex}^{0}(q_x,q_y)$ (upper) and SC susceptibility $\psi_{\rm sc}^{0}(q_x,q_y)$ (lower) in $q_x$-$q_y$ plane for (a) $D=-0.50$ $(T=0.030)$ and (b) $D=-0.65$ $(T=0.015)$, respectively. 
}
\label{Fig9}
\end{figure}
The present phase diagrams for normal-EP and normal-SC phase 
are attributed to the intrinsic $c$-$f$ density imbalance of the present quasi 1-D property 
which yields two types of the $c$-$f$ nesting effects 
between the $c_{-}$-$f$ and $c_{+}$-$f$ bands 
in the particle-hole (excitonic) and particle-particle (superconducting) susceptibilities 
$\chi_{\rm ex}^{0}(\bm{q})$ and $\psi^{0}_{\rm sc}(\bm{q})$. 
In order to see such the quasi 1-D structure explicitly, 
we plot the $\bm{q}$-dependence of $\chi_{\rm ex}^{0}(\bm{q})$ and $\psi^{0}_{\rm sc}(\bm{q})$ 
in $q_x$-$q_y$ plane for several values of $D$ and $T$ in Fig. \ref{Fig9}
where (a) $D=-0.50$ $(T=0.030)$ and (b) $D=-0.65$ $(T=0.015)$, respectively. 

Both of $\chi_{\rm ex}^{0}(q_x,q_y)$ and $\psi_{\rm sc}^{0}(q_x,q_y)$ 
shows the clear quasi 1-D peak structures as a function of $q_x$ and their weak $q_y$-dependences 
over the $c$-$f$ compensated semimetallic region with three-FSs 
as shown upper and lower lines of Fig. \ref{Fig9}, respectively. 
With decreasing $D$ and $T$, 
the both $q_x$ giving the peak of $\chi_{\rm ex}^{0}(q_x,q_y)$ and $\psi_{\rm sc}^{0}(q_x,q_y)$ increase, 
where $q_x=0.125\pi\sim 0.2\pi$ of $\chi_{\rm ex}^{0}(q_x,q_y)$ 
corresponds to the EP instabilities towards from FFLO1 EP to FFLO2 EP 
and $q_x=0.4\pi\sim 0.45\pi$ of $\psi_{\rm sc}^{0}(q_x,q_y)$ 
corresponds to the SC phase instabilities towards from FFLO2 SC phase to FFLO3 SC phase. 
Such the various structures of $\chi_{\rm ex}^{0}(q_x,q_y)$ and $\psi_{\rm sc}^{0}(q_x,q_y)$ in $q_x$-$q_y$ plane 
is attributed to the newly-added inter-chain hoppings which yields the different curvature of FSs between $c_{+}$ and $c_{-}$-bands.

\end{document}